\documentclass[useAMS,usenatbib]{mn2e}

\usepackage{graphicx}

\title[Covert connection of filaments]
{Covert connection of filaments}
\author[B. Filippov]{B. Filippov \thanks{E-mail:
bfilip@izmiran.ru}  \\ Pushkov Institute of Terrestrial Magnetism,
Ionosphere and Radio Wave Propagation of the Russian Academy of
Sciences (IZMIRAN), \\ Troitsk, Moscow 142190, Russia}

\begin{document}

\date{Accepted 0000 December 15. Received 0000 December 14; in original form 0000 October 11}

\pagerange{\pageref{firstpage}--\pageref{lastpage}} \pubyear{2002}

\maketitle

\label{firstpage}

\begin{abstract}
We analyse the relationship between two near filaments, which do
not show any connection in H$\alpha$  images but reveal close
magnetic connectivity during filament activations in Extreme
Ultraviolet (EUV) observations.  A twisted flux rope, which
connects a half of one filament with another filament, becomes
visible during several activations but seems to exist all the time
of the filaments presence on the disc. {\bf {\it Solar Dynamic
Observatory} ({\it SDO}) and {\it Solar Terrestrial Relations
Observatory} ({\it STEREO}) observed the region with the filaments
from two points of view separated by the angle of about
120$^\circ$. On 2012 July 27, {\it SDO} observed the filament
activation on disc, while for the {\it STEREO B} position the
filaments were visible at the limb.  Nearly identical interaction
episode was observed on 2012 August 04 by {\it STEREO A} on disc
and by {\it SDO} at the limb. This good opportunity allows us to
disentangle the 3-D shape of the connecting flux rope and in
particular to determine with high reliability the helicity sign of
the flux rope, which looks ambiguous in preliminary inspections of
on-disc EUV images only.} The complex magnetic structure of the
region consists of three braided flux ropes in the vicinity of the
coronal null point. Using observations of the flux rope fine
structure and plasma motions within it from two points of view, we
determine the negative sign of helicity of the flux rope, which
corresponds to dextral chirality of the filaments. The
observations, despite the tangled fine structure in some EUV
images, support flux rope filament models. They give more evidence
for the one-to-one relationship between the filament chirality and
the flux rope helicity.
\end{abstract}


\section{Introduction}

Solar filaments, or prominences as they are called when observed
above the limb, are cool plasma formations supported against the
gravity by magnetic field high in the corona. Two magnetic
configurations are most often considered as a magnetic skeleton of
filaments. They are a twisted magnetic flux rope \citep{b63, b46,
b45, b71} and a sheared arcade \citep{b60, b68, b2, b13, b49, b3,
b33, b47} . Both of them have some observational bases and both
meet with certain difficulties. Flux ropes become most evident
during filament activations and eruptions. However, some authors
raise objections against existence of helical structures within
filaments before eruptions. They suggest that twisted flux ropes
are forming during the eruptions after field line reconnection
\citep{b24, b1, b51}. One of the vague points is how flux ropes
appear in the corona. Do they emerge form the convection zone
being formed there by convection and differential rotation
\citep{b56}, or they are created in the corona from potential
fields due to photospheric footpoint motions and reconnection
\citep{b54, b43, b25} ?

\begin{figure*}
\includegraphics[width=167mm]{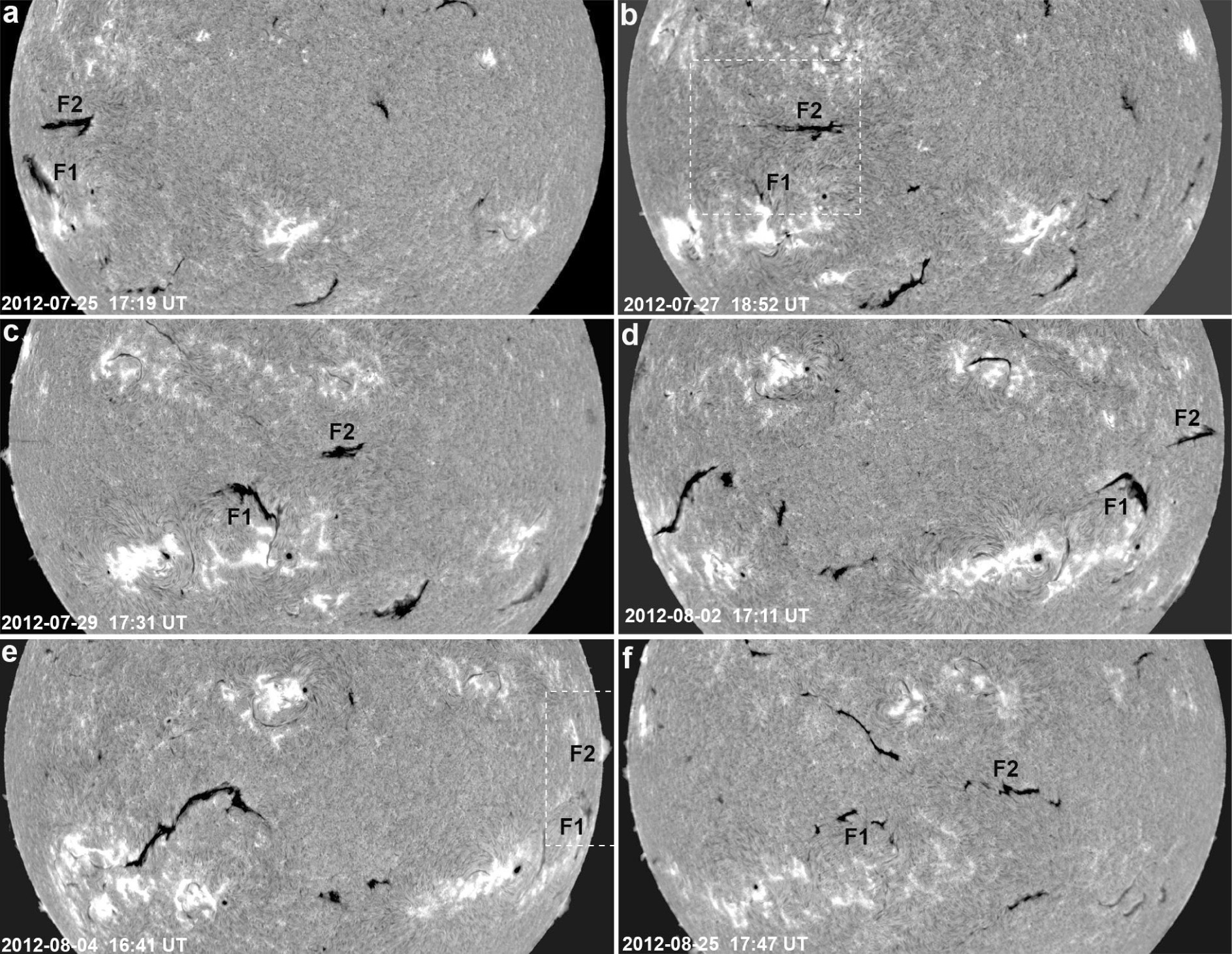}
\caption{H$\alpha$  filtergrams showing the filaments F1 and F2
during their passage through the solar disc and on the next
rotation. White dashed squares correspond to the field-of-view of
the images in Fig. 2 and 4. (Courtesy of the Big Bear Solar
Observatory). }
\end{figure*}

Recently, many observations of flux ropes in the corona were
reported \citep{b61, b9, b10, b72, b39,b40, b41, b58}. Low optical
thickness of coronal structures in Extreme Ultraviolet (EUV) makes
it not so easy to derive the true helicity sign of observed
helical structures. In many cases, it is difficult to decide which
features are closer to an observer and which are farther. With
flux ropes, we cannot be sure which threads are on the upper side
of them and which threads belong to the bottom side. A wrong
choice leads to the wrong estimation of the helicity sign.
Simultaneous observations from different points of view provided
by {\it Solar Dynamic Observatory} ({\it SDO}), and {\it Solar
Terrestrial Relations Observatory} ({\it STEREO}) give good
opportunity to reconstruct the 3-D shape of coronal flux ropes.

In active regions, filaments consist of long thin threads
stretched along polarity inversion lines (PIL). This hints on the
presence of a strong axial magnetic-field component. Indeed,
measurements of the magnetic field in prominences showed that the
magnetic field in prominences is mostly horizontal and makes an
acute angle of about 20 - 40$^\circ$ with respect to the long axis
of the prominence \citep{b38, b11, b6}. According to the direction
of the axial component of the filament magnetic field relative the
surrounding photospheric fields, all filaments can be divided into
two classes. A filament is called 'dextral' if the axial component
is directed toward the right when the filament is viewed from the
side of the positive photospheric polarity, and 'sinistral' if the
direction of the axial component is opposite to this \citep{b52}.
Usually on both sides of a filament, filament barbs are observed
protruding at an acute angle from the main body of the filament.
They can be classified as either right-bearing or left-bearing
depending on the deviation of the barbs from the axis. In most
cases, the barbs of a dextral/sinistral filament are observed to
be right/left bearing \citep{b50}. Thin threads composing the main
filament body are deviated at an acute angle clockwise to the axis
in dextral and counterclockwise in sinistral filaments. This makes
it possible to determine the class of a filament (the filament
chirality) from its visual appearance, without information on the
magnetic fields \citep{b59}.

\begin{figure*}
\includegraphics[width=167mm]{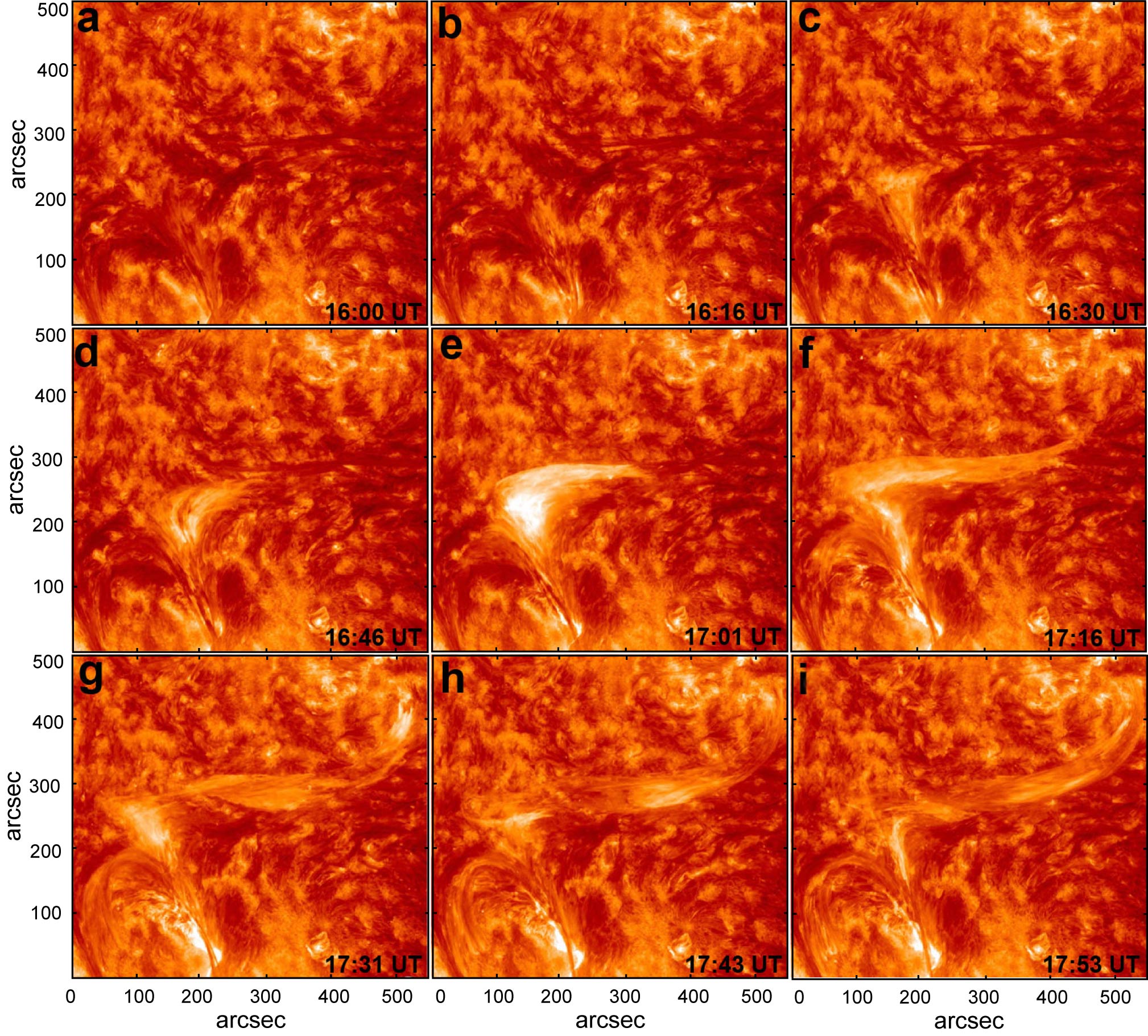}
\caption{Filament activation observed on 2012 July 27 by {\it
SDO}/AIA in the 304-\AA \ channel on disc. (Courtesy of the
NASA/{\it SDO} and the AIA science team.)}
\end{figure*}

\begin{figure*}
\includegraphics[width=167mm]{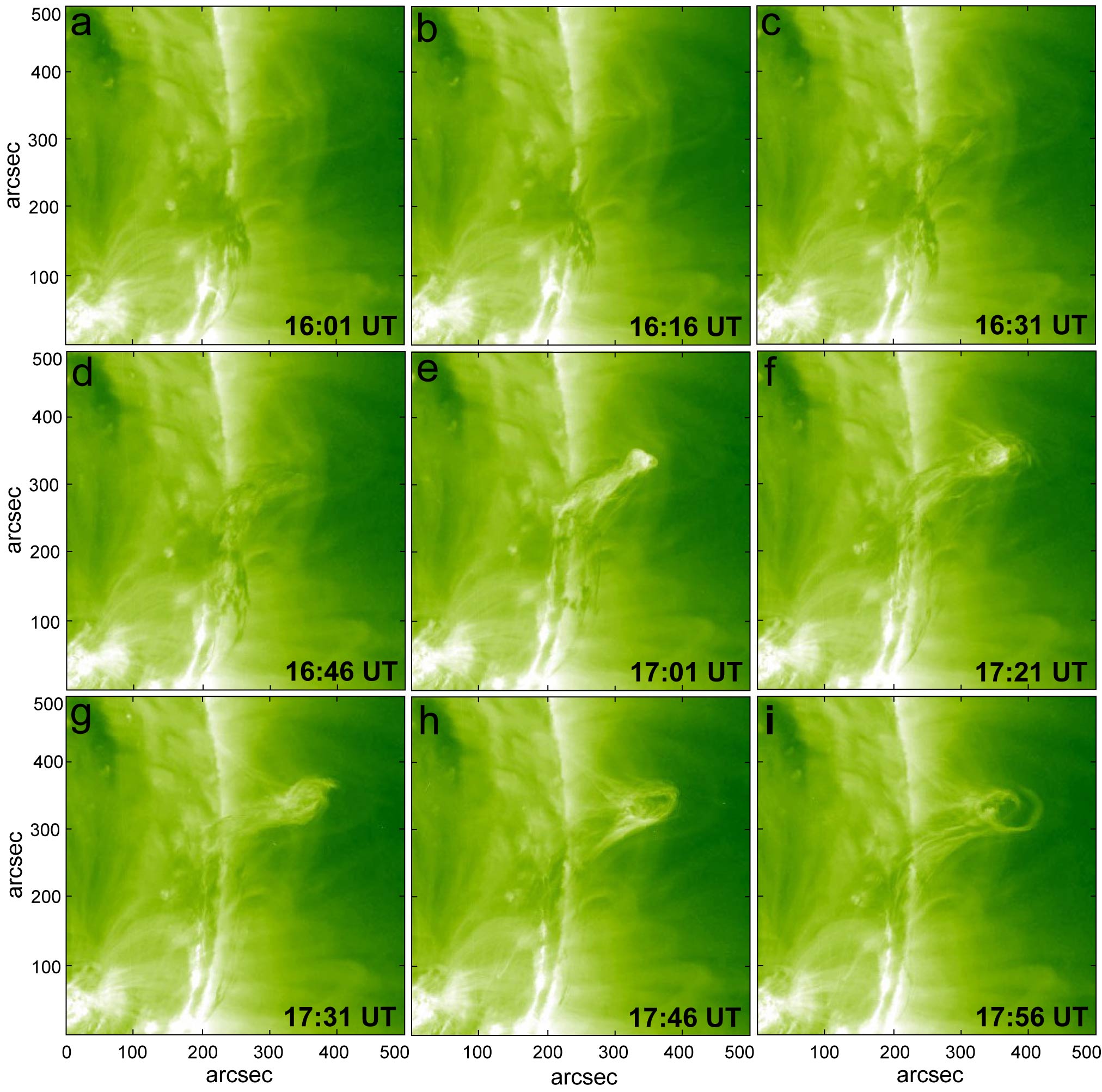}
\caption{Filament activation observed on 2012 July 27 by {\it
STEREO B}/SECCHI in the 195-\AA \ channel at the limb. (Courtesy
of the {\it STEREO}/SECCHI Consortium.)}
\end{figure*}

\begin{figure*}
\includegraphics[width=167mm]{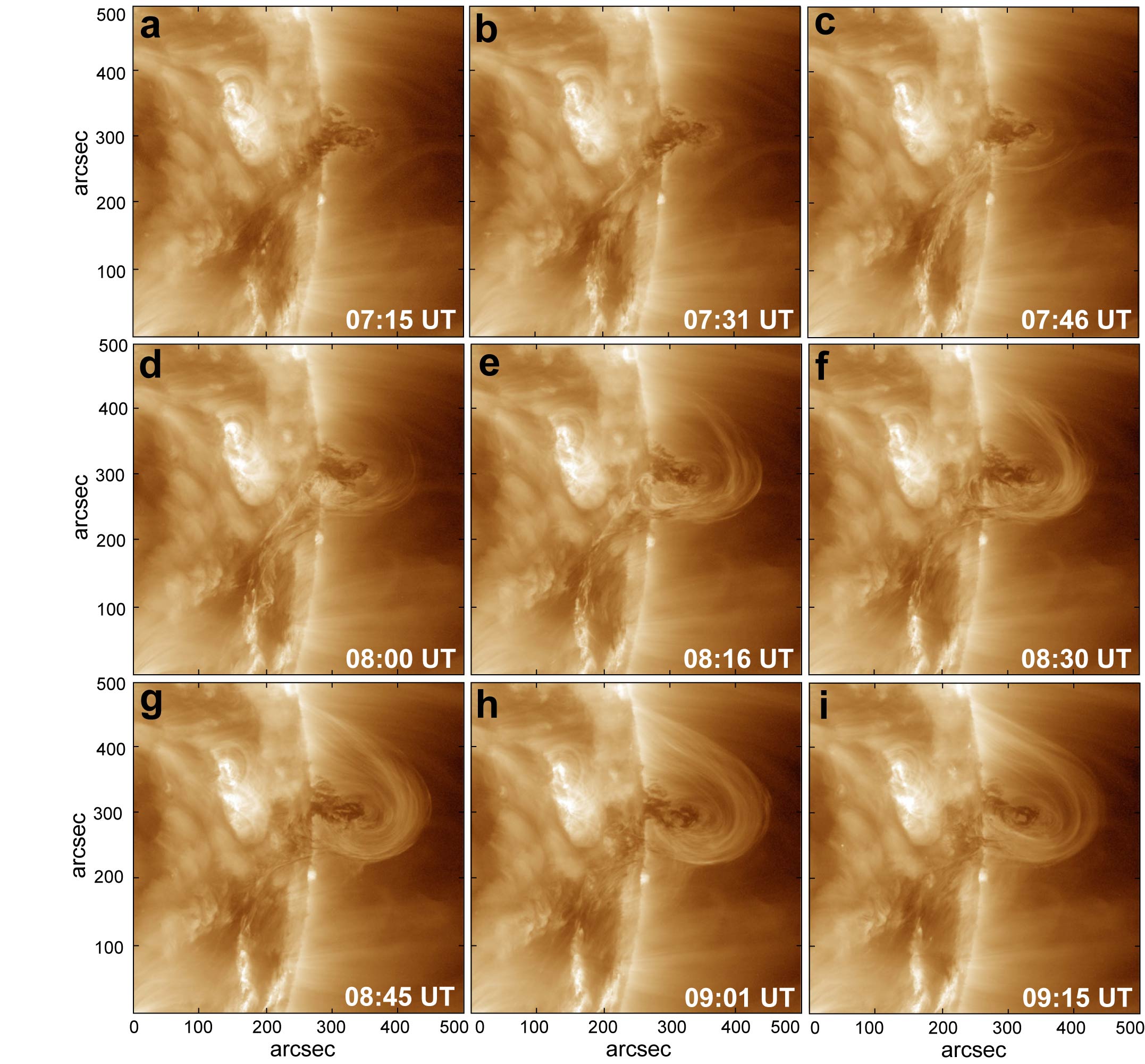}
\caption{Filament activation observed on 2012 August 04 by {\it
SDO}/AIA in the 193-\AA \ channel at the limb. (Courtesy of the
NASA/{\it SDO} and the AIA science team.) }
\end{figure*}

\begin{figure*}
\includegraphics[width=167mm]{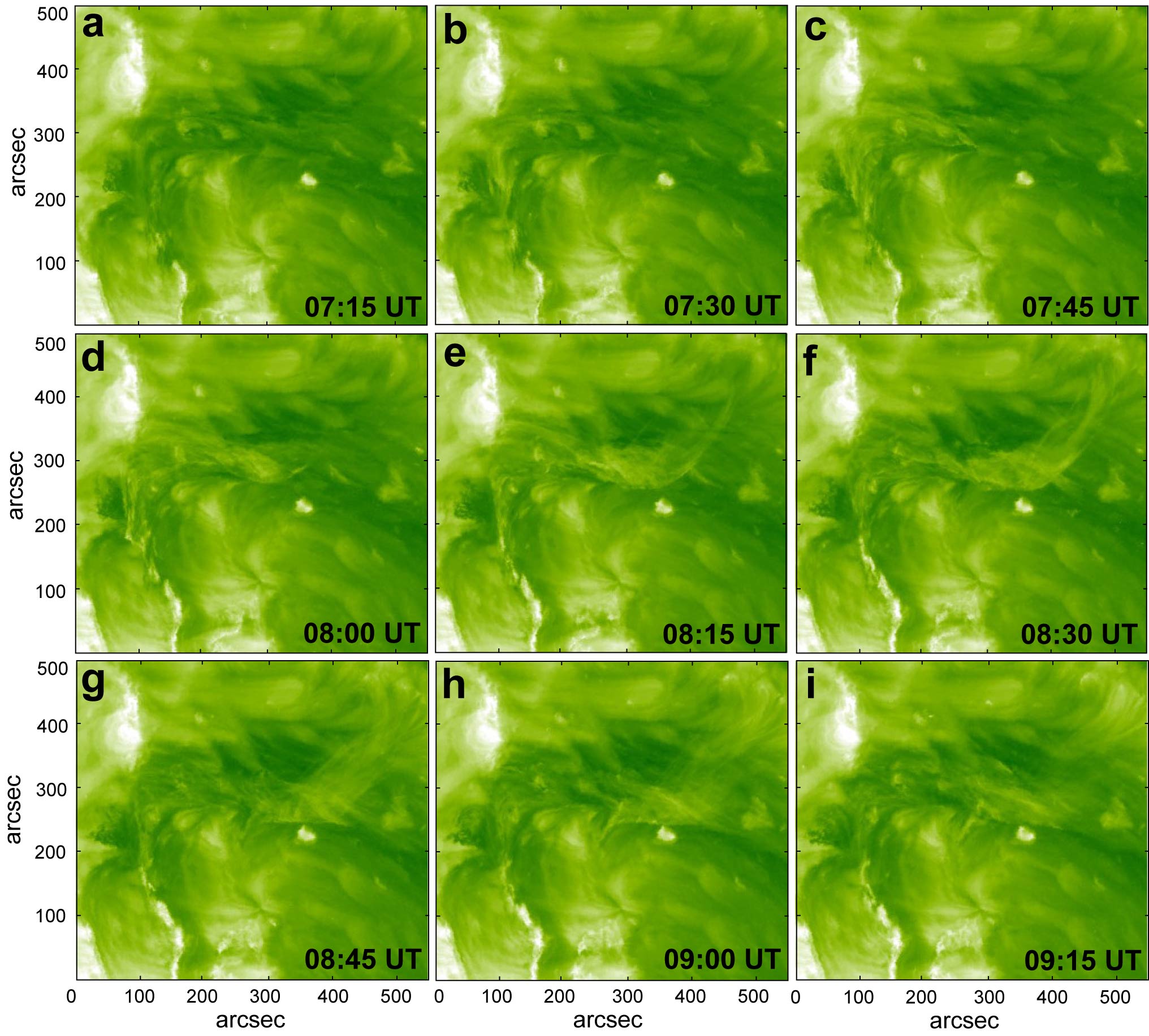}
\caption{Filament activation observed on 2012 August 04 by {\it
STEREO A}/SECCHI in the 195-\AA \ channel on disc. (Courtesy of
the {\it STEREO}/SECCHI Consortium.)}
\end{figure*}

The bottom parts of helical flux tubes serve as potential wells or
magnetic traps where dense and cold plasma can be collected
forming the fine structure of the filaments observed in
chromospheric lines. During a filament activation or eruption,
plasma that moves along field lines can spill over the upper parts
of the flux tubes and show both the bottom and upper parts of
helices. Recent observations strongly support the idea that
dextral filaments are associated with left-handed flux ropes with
negative helicity, while sinistral filaments are associated with
right-handed flux ropes with positive helicity \citep{b62, b7,
b32}.

It was well known for a long time that solar filaments follow
large scale PILs of the photospheric magnetic field \citep{b4,
b26, b66, b53, b67, b14, b28}. If two PILs are located not far
from each other, they can approach closer due to evolution of the
photospheric field and change connectivity ('reconnect') after
their contact at some point. A pair of filaments of the same
chirality may exchange by their halves and also reconnect, while a
pair of filaments of the opposite chirality breaks into four
independent filaments \citep{b36, b8a, b67a, b16, b21, b17}. When
two long filaments cross each other to form a cruciform structure,
two intersecting PILs correspond to a quadrupole magnetic
configuration. There is a null point near the site of intersection
in the magnetic field created by photospheric sources.

Merging of short filament segments of the same chirality aligned
along the same PIL into a single structure was studied by
\citet{b64a}. Two filaments of opposite chirality  in the same
decaying active region interacted without merging but produced a
confined flare \citep{b12a}.

In this paper, we analyze the relationship between two near
filaments, which do not show any connection in H$\alpha$  images
but reveal the close magnetic connectivity during filament
activations in EUV channels. A twisted flux rope, which connects a
half of one filament with another filament, becomes visible during
several activations but seems to exist all the time of the
filaments presence on the disc. Using observations of the flux
rope structure and plasma motions within it from two points of
view, we determine the negative sign of helicity of the flux rope,
which corresponds to dextral chirality of the filaments.

\section[]{Data and General Description of Events}

We used observations of the Big Bear Solar Observatory, {\it
STEREO}, and {\it SDO}. Synoptic full-disk H$\alpha$  images
(Fig.1) in the end of July and in the beginning of August 2012
show two filaments located not far from each other at a middle
latitude in the southern hemisphere. They are marked as F1 and F2
in Fig. 1. The filaments are well developed and quite stable. They
do not change significantly their shapes during the passage
through the solar disc in the end of July and in the beginning of
August and can be easily identified on the next solar rotation in
the end of August. The filaments seem to be not connected with
each other in images in H$\alpha$  line. However, observations in
EUV with the Sun Earth Connection Coronal and Heliospheric
Investigation (SECCHI) Extreme Ultraviolet Imager (EUVI;
\citealt{b70, b27}) on board {\it STEREO} and with the Atmospheric
Imaging Assembly (AIA; \citealt{b37}) on board the {\it SDO}
(Figs. 2 - 5) show that during activations the filaments reveal
the connectivity of their magnetic structures.

Several pulses of filament activations were observed between July
27 and August 04. The most dramatic events happened on July 27
between 16 and 18 UT, on July 29 between 00:30 and 02:30 UT, and
on August 04 between 07 and 10 UT. Each of them starts from
brightening at the south-western end of the horseshoe-shaped
filament F1. Bright threads elongate along the filament axis
showing field aligned motion of hotter plasma. Figs. 2, 3 show the
third activation on July 27 observed by {\it SDO}/AIA in the
304-\AA \ channel on disc and by {\it STEREO B}/SECCHI in the
195-\AA \ channel at the limb. The separation angle of the {\it
STEREO} spacecraft with the Earth was about 120$^\circ$ in summer
2012. Long thin threads reveal the axial magnetic field in the
filament channel. However, most of threads not follow the
horseshoe-curved axis of the filament F1 but make nearly
right-angle turn to the west and follow the axis of the filament
F2. The eastern part of the filament F1 becomes invisible in
H$\alpha$  images during activations (Fig. 1(b)).

Just the same scenario was observed on 04 August, only the aspect
angles for the spacecrafts changed to opposite. {\it SDO} observed
the filaments close to the limb, while {\it STEREO A} observed
events on disc (Figs. 4, 5). Better time and spatial resolution of
{\it SDO}/AIA images allow recognizing finer details in the
structure of the filaments both on disc on July 27 and at the limb
on August 04. Since the events on both days are very similar, we
can expect that the magnetic structure of the filaments has not
changed in this period of time and may consider all images as
different views of the same structure. Thus, the filaments that
look separate and independent in H$\alpha$  images are contained
within a united magnetic structure.

\begin{figure}
\includegraphics[width=84mm]{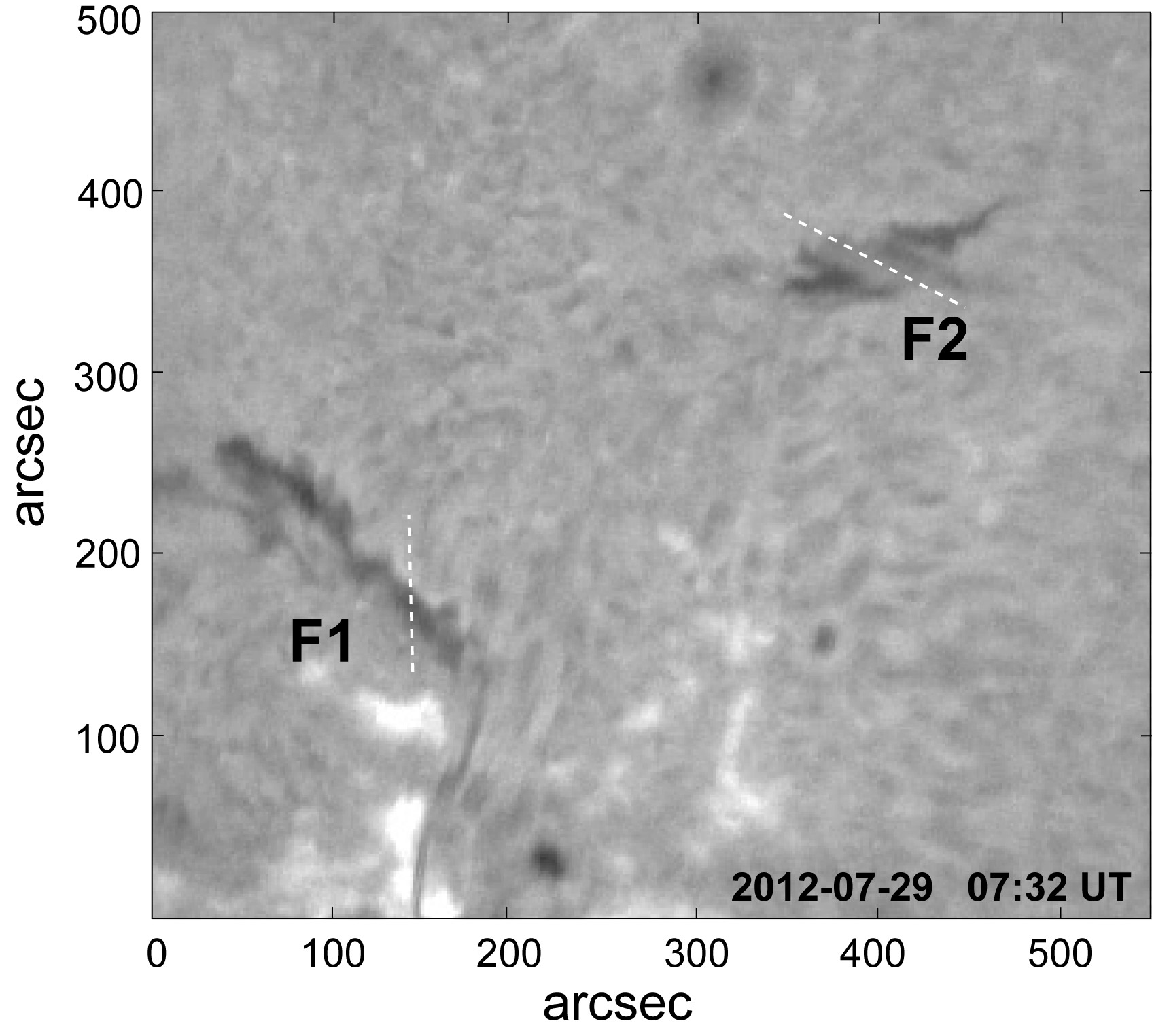}
\caption{H$\alpha$  spectroheliogram showing the chirality of the
filaments F1 and F2. White dashed lines show the direction of
filament threads deviating  at an acute angle clockwise to the
axes of filaments (Courtesy of the Observatory Paris-Meudon). }
\end{figure}

Despite the location in the southern hemisphere, the chirality of
the filament F1 is dextral. This is very evident from the fine
structure of the filament in H$\alpha$  line (Figs. 1 (a), (c),
(d), 6). Thin threads within the filament body deviate clockwise
from the filament axis (Fig. 6) and filament barbs are right
bearing, which both correspond to the dextral chirality. According
to one-to-one correspondence between the filament chirality and
the enveloping flux rope helicity \citep{b63, b62, b7, b59, b48,
b8, b32} we may expect the manifestation of a left-handed flux
rope during filament activations. However some images look
puzzling.

\begin{figure*}
\includegraphics[width=167mm]{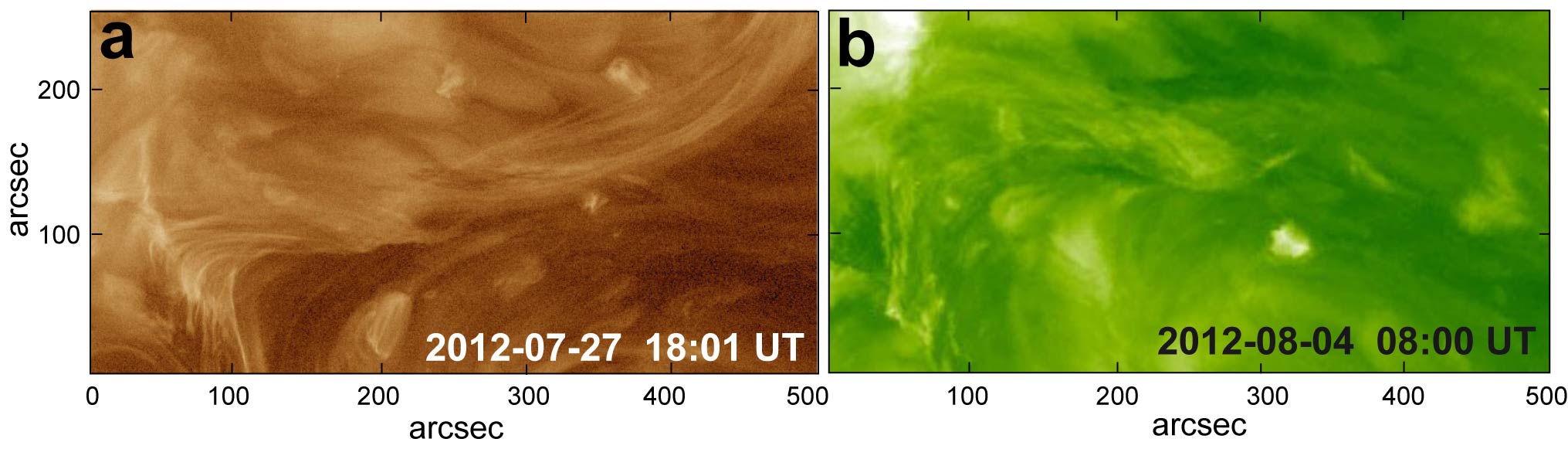}
\caption{Flux-rope structure looking like a right-hand helix
observed on 2012 July 27 by {\it SDO}/AIA in the 193-\AA \ channel
(a) and on 2012 August 04 by {\it STEREO A}/SECCHI in the 195-\AA
\ channel (b). (Courtesy of the NASA/{\it SDO} and the AIA science
team and {\it STEREO}/SECCHI Consortium.) }
\end{figure*}

\section[]{Puzzling Structures}

\begin{figure*}
\includegraphics[width=140mm]{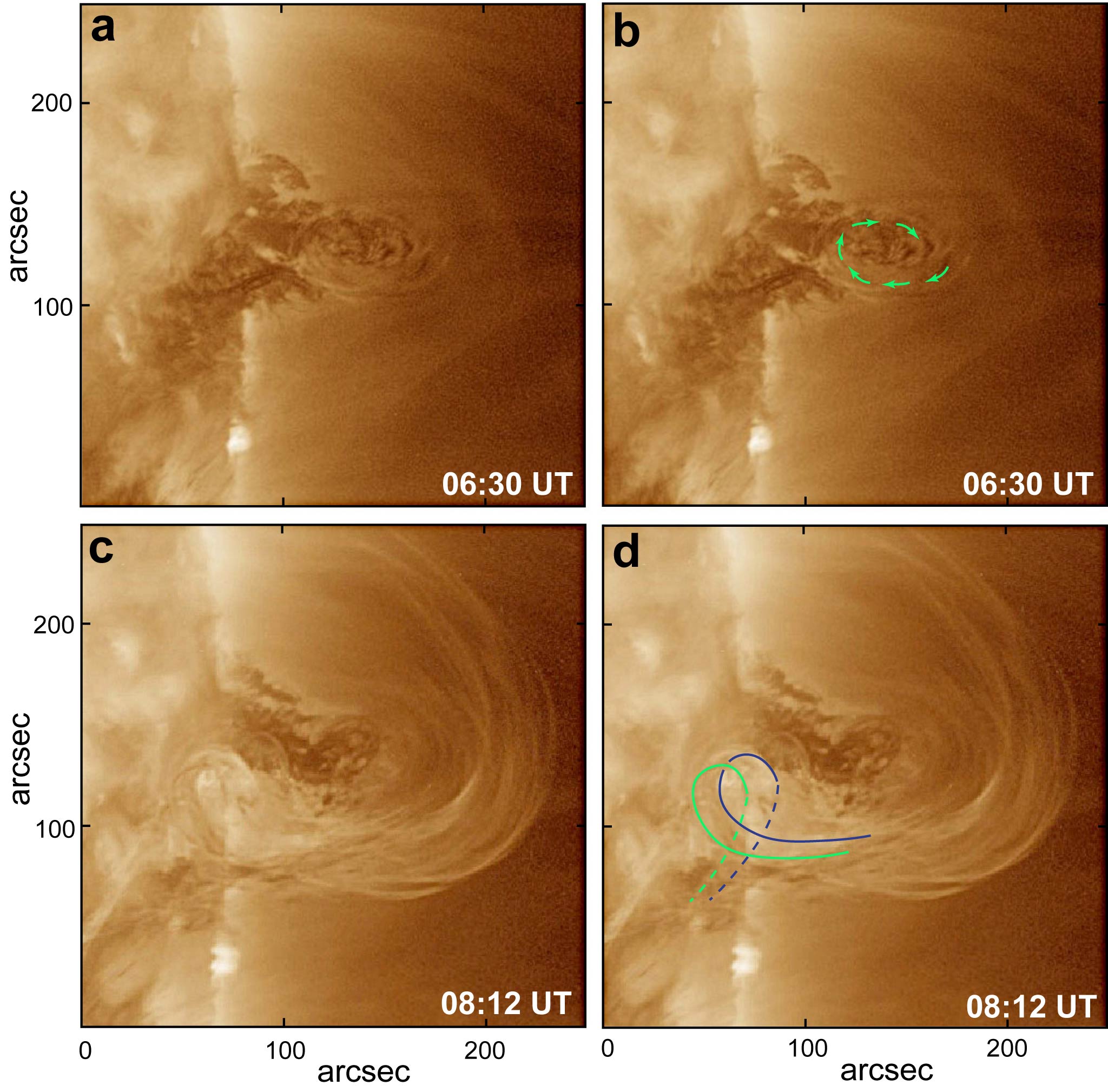}
\caption{Clockwise rotation of elongated blobs within the body of
the filament F2 (a), (b) and a loop-like structure at the lower
part of the flux rope (c), (d) observed by {\it SDO}/AIA in the
193-\AA \ channel. (Courtesy of the NASA/{\it SDO} and the AIA
science team.)}
\end{figure*}

\begin{figure}
\includegraphics[width=84mm]{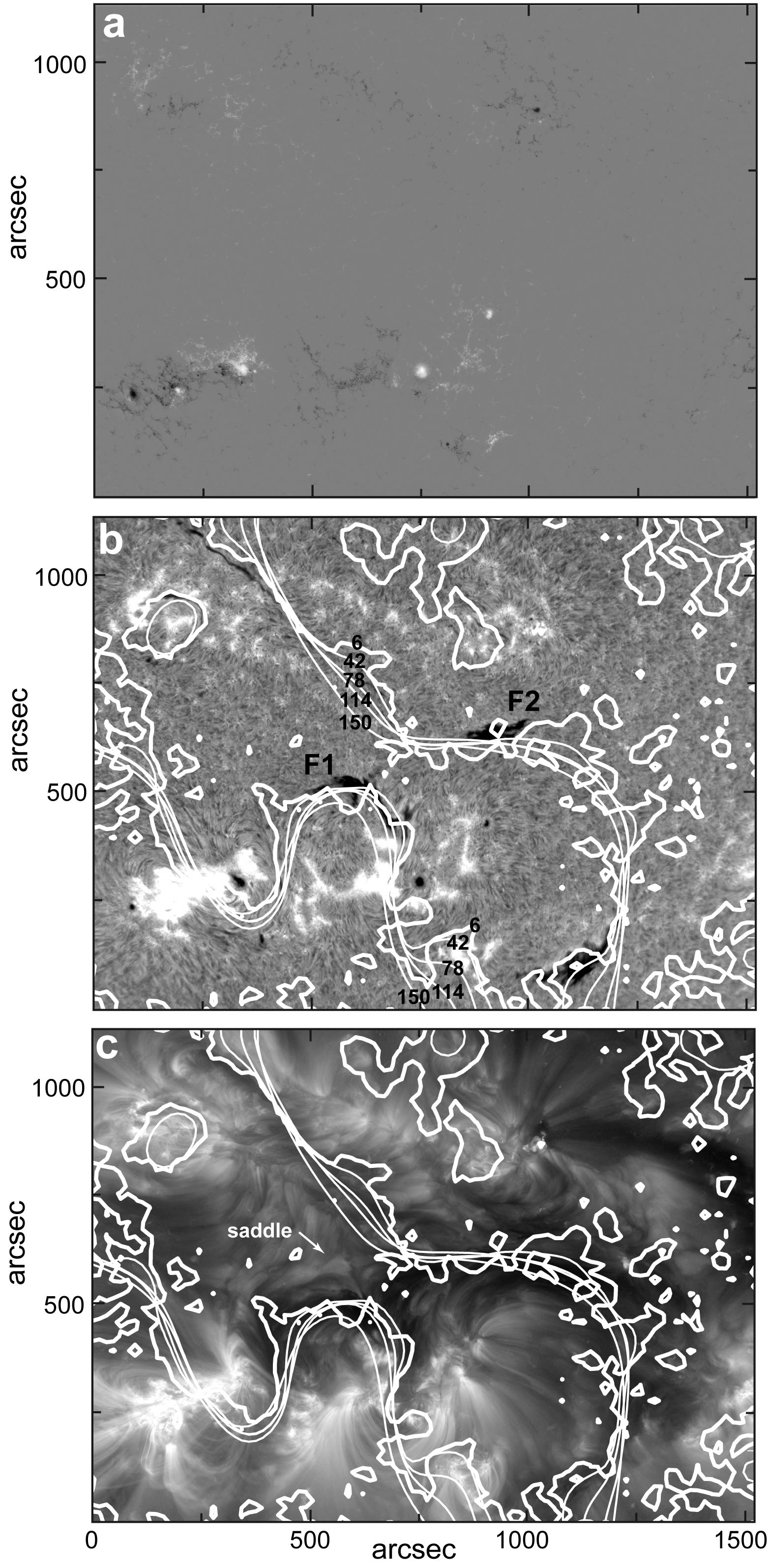}
\caption{Fragment of the HMI magnetogram taken on 2012 July 29 at
17:31 UT containing all most important magnetic sources on the
disc (a), H$\alpha$  filtergram of the same region and at the same
time with superposed PILs at a height of 6 (thick lines), 42, 78,
114, and 150 Mm (b), {\it SDO}/AIA image in the 193-\AA \ channel
with superposed PILs (c). (Courtesy of the NASA/{\it SDO} HMI and
AIA science teams and of the Big Bear Solar Observatory.)}
\end{figure}

\begin{figure*}
\centering
\includegraphics[width=167mm]{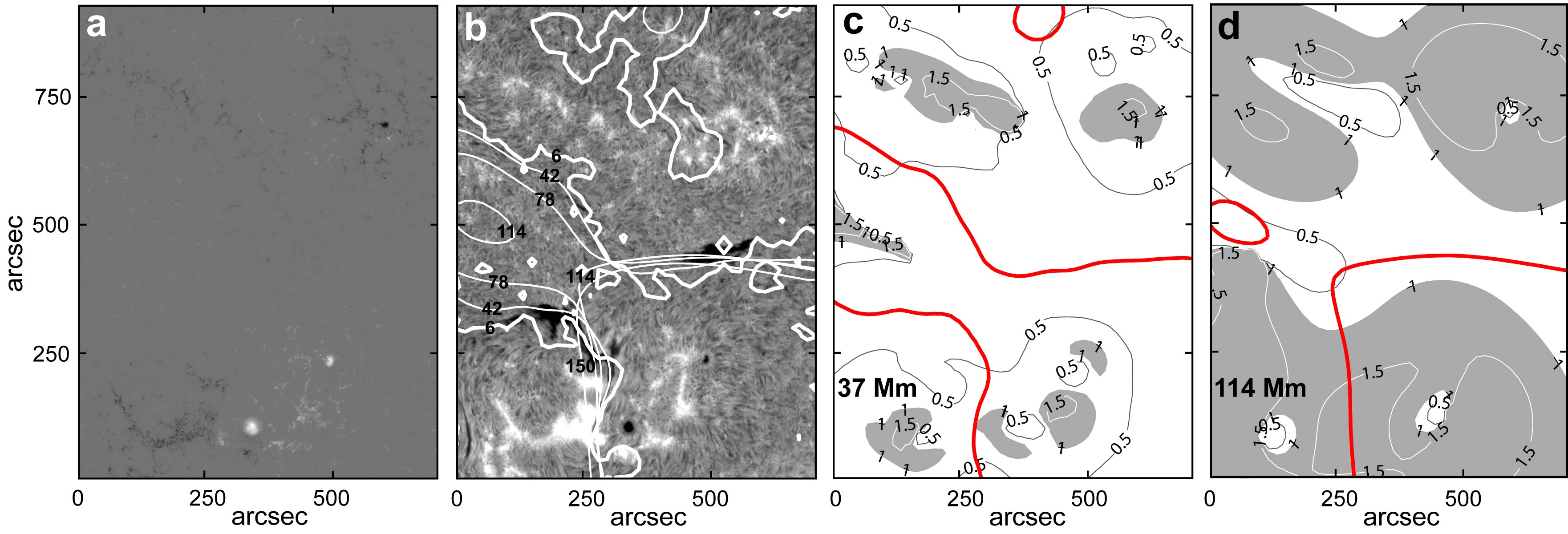}
\caption{Smaller fragment of the HMI magnetogram taken on 2012
July 29 at 17:31 UT (a), H$\alpha$  filtergram of the same region
and at the same time with superposed PILs (b), and distributions
of the decay index and PILs (thick red lines) at different heights
(c) - (d). Shadowed areas show the regions where $n > 1$.
(Courtesy of the NASA/{\it SDO} and the HMI science team and of
the Big Bear Solar Observatory.)}
\end{figure*}

Beside the covert connection of the filaments F1 and F2 revealed
during several activations, some EUV images look puzzling if we
keep in mind the expected correspondence between the filament
dextral chirality and the negative helicity of the enveloping flux
rope. Fig. 7 shows the flux-rope structure looking like a
right-hand helix on 2012 July 27 in the {\it SDO}/AIA image in the
193-\AA \ channel and on 2012 August 04 in the {\it STEREO
A}/SECCHI image in the 195-\AA \ channel (see also Fig. 2(i) and
Fig. 5(c) - (e)). Most of thin bright threads forming the flux
rope deviate clockwise from the flux-rope axis and only few faint
threads are visible deviating counterclockwise. It is unknown for
certain which threads are higher and which threads are lower but
one can expect that the higher threads would be more noticeable in
images. Thus, the structures in Fig. 7 give us impression of a
right-hand helix.

Another puzzling feature is a clockwise rotation of elongated
blobs within the filament F2 observed by {\it SDO}/AIA at the limb
between 06:15 UT and 07:30 UT (Fig. 8(a), (b), movie\_SDO\_04). If
activation starts at the south-western end of the filament F1 and
propagates from the east to the west along the filament F2, we
could expect a counterclockwise rotation of blobs moving along a
left-handed helix away from us. Instead, the rotation is clockwise
during more than an hour.

One more puzzle is a loop-like structure at the lower part of the
flux rope (Fig. 8(c), (d)). The ascending branch of the loop seems
to be located in front of other threads. Its continuation runs
above the dark body of the filament F2 to the north-west (Fig.
5(d) - (g)). The other ends of the loop-like threads represented
by dashed lines in Fig. 8(d) should pass below the ascending
branch to the location of the filament F1. Therefore, the most
curved segments of the lines correspond to a 180$^\circ$ turn of
the threads and show a real loop. There is no evidence of such a
loop in {\it STEREO A}/SECCHI images in a view from above (Fig.
5).

In order to recognize the true geometry of the observed structures
we need more detailed and careful examination of all images and
correlation of them with a presumable magnetic configuration.

\begin{figure*}
\includegraphics[width=167mm]{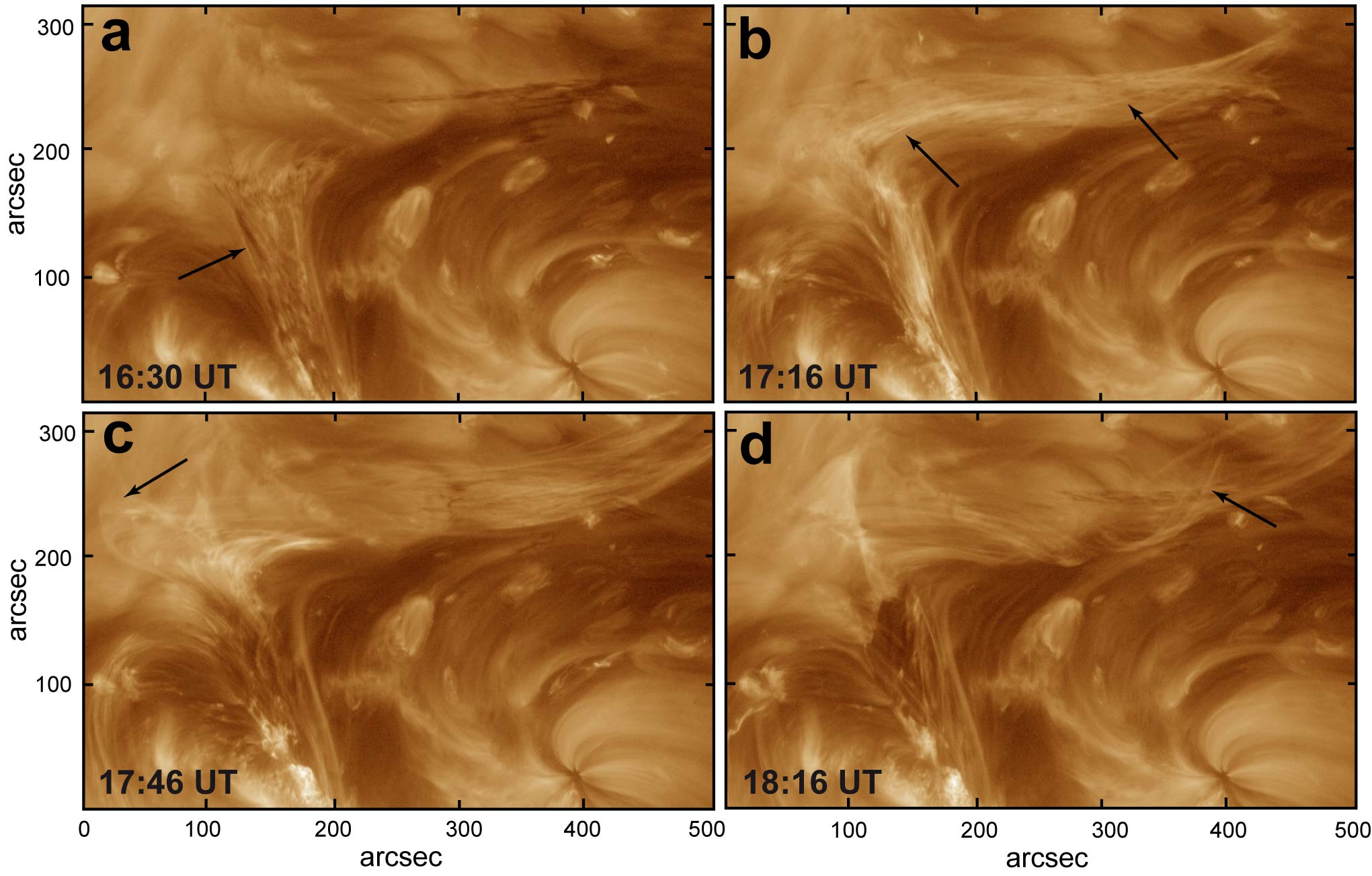}
\caption{Filament structure observed on 2012 July 27 by {\it
SDO}/AIA in the 193-\AA \ channel, which show features undoubtedly
corresponded to left-hand helix (a), (b), (d) and very curved
threads above the saddle (c). (Courtesy of the NASA/{\it SDO} and
the AIA science team.)}
\end{figure*}

\begin{figure*}
\includegraphics[width=167mm]{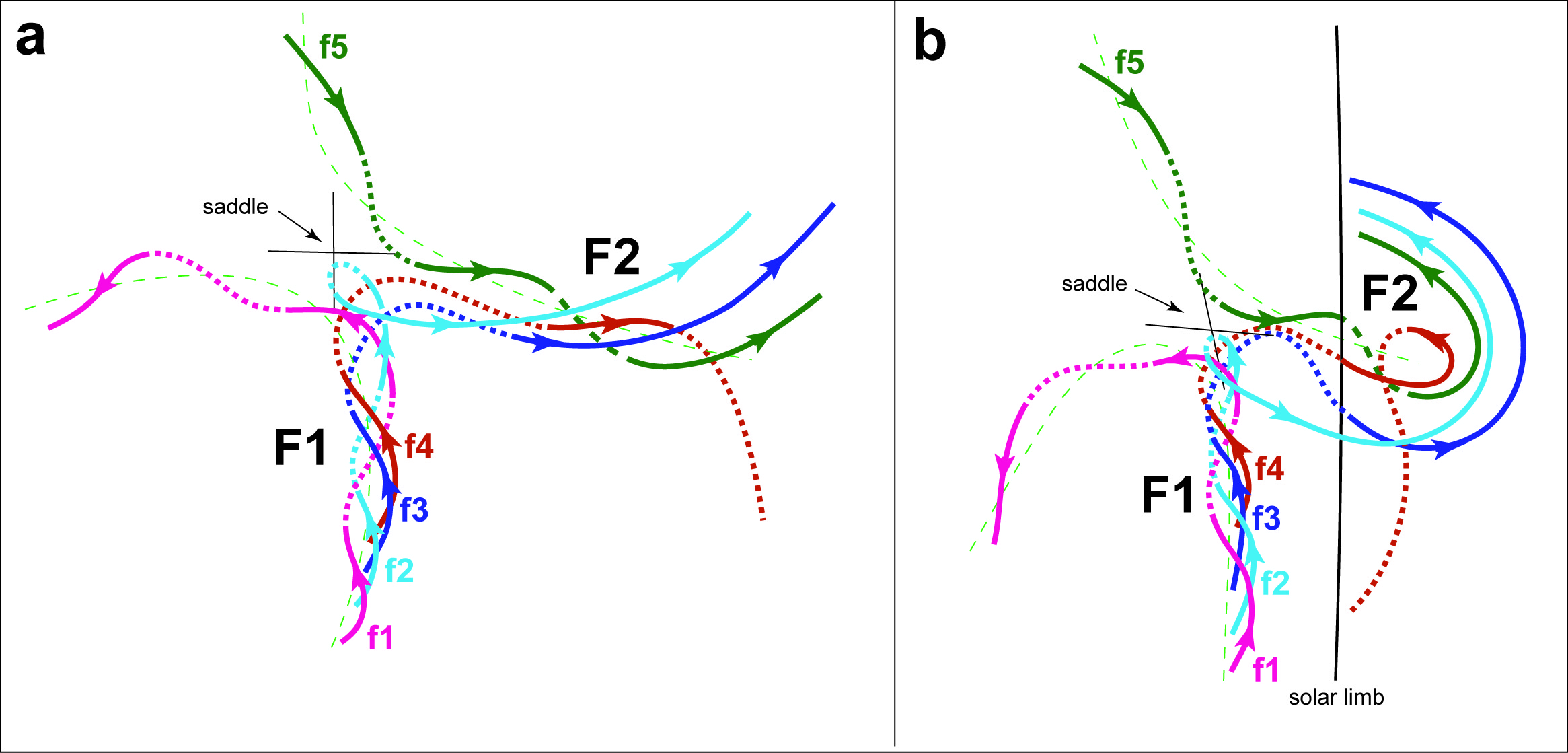}
\caption{Scheme of the magnetic field structure of two filaments.
Top view (a) and side view (b).}
\end{figure*}

\section[]{Magnetic Configuration}

When two long filaments cross each other to form a cruciform
structure, two intersecting PILs correspond to a quadrupole
magnetic configuration. Our filaments F1 and F2 do not approach
very close to each other and do not demonstrate any connection in
a quiescent state. However every activation reveals the field
lines that connect them.

\subsection[]{Polarity Inversion Lines of Potential Magnetic Field}

Fig. 9(a) shows a fragment of the magnetogram taken by the
Heliospheric and Magnetic Imager (HMI; \citealt{b65}) on board
{\it SDO} on 2012 July 29 at 17:31 UT when the filaments F1 and F2
were close to the central meridian (Fig. 1(c)). The fragment
contains all most important magnetic sources on the disc. Using
this magnetogram as the boundary condition for the potential
magnetic field calculations we calculate $B_z$ maps ($z$-axis is
vertical) at different heights and obtain a set of PILs $B_z = 0$.
We solve the Neuman external boundary value problem numerically
using the Green function method \citep{b19, b20, b18}. Our
numerical code of a potential field extrapolation is more suitable
for rather small solar areas because it assumes the considered
photospheric area as a flat surface. The fragment in Fig. 9(a)
covers a large part of the disc and does not fit to this
condition. Therefore we will consider the results of calculations
as a rough approximation.

In Figs. 9(b), (c) the PILs are superposed on the H$\alpha$
filtergram and on the 193-\AA \ filtergram. The lowest PIL at a
height of 6 Mm is shown by a thicker line. The higher PILs do not
approach closer as it sometimes happens in quadrupolar
configurations \citep{b21, b16}. However a saddle-like structure
is visible near the closest approach of the PILs (Fig. 9(d)). It
manifests the real presence of the quadrupolar configuration.

If we cut a smaller fragment from the same HMI magnetogram (Fig.
10(a)), the result is different. At the height of about 100 Mm,
PILs running above the filament F1 turn to the west and run above
the filament F2 (Fig. 10(b)). At these heights, a flux rope can
exist, which is able to contain both the south-western part of the
filament F1 and the filament F2. The flux rope is in horizontal
equilibrium on the PIL. Stable vertical equilibrium is possible
only if the so-called decay index $n$ = - dln$B$/dln$h$ of the
surrounding magnetic field is less than the critical value $n_c$
\citep{b69, b5, b23, b42, b34, b15, b31, b12, b57, b55}. The exact
value of $n_c$ depends on parameters of a flux-rope model and
varies between 1 and 1.5 \citep{b69, b5, b19, b20, b12}. Figs.
10(c)-(d) show the distribution of the decay index at heights of
37 and 114 Mm. At the height of 37 Mm, the shadowed area $n > 1$
first touches the PIL near the south-western end of the filament
F1. Possibly, it is connected with the initiation of all
activations in this place. The area around the filament F2 is
stable up to the heights much above 100 Mm.

The results shown in Fig. 10 are also a rough approximation. On
the one hand, the boundary area is smaller and closer to the flat
surface. On the other hand, some strong magnetic sources are not
taken into account. Particularly, these are active regions to the
east from the selected area. They 'attract' the PIL and do not
allow it to turn to the west in Fig. 9 like it does at high
heights in Fig. 10. Despite of their roughness, the results of
potential magnetic field calculations confirm the presence of the
quadrupolar magnetic configuration at the place where the
filaments approach each other and show the possibility of the
existence of the stable flux rope connecting the filaments F1 and
F2.

\subsection[]{Fine Structure of the Flux Rope}

It is most probable that flux tubes connecting the vicinities of
F1 and F2 exist at all times during their passage through the
solar disc in the end of July and in the beginning of August. They
become visible after F1 activations when they are filled with both
hot and cold plasmas. Evidently they are twisted into a flux rope.
This flux rope can consist of some strands of the flux ropes
containing the main bodies of the filaments F1 and F2 or be a
separate flux rope like in double-decker filaments \citep{b44,
b35}.

Coronal images give us 2-D projections of real 3-D coronal
structures. Sometimes, it is difficult to derive the true
geometrical shape of observed formations from only one projection.
In a case of flux ropes, it may lead to estimation of wrong sign
of helicity if we do not know whether observed features belong to
the frontal or far side of the flux rope. Observations from
another view point and internal motions within the formation allow
us to derive more correctly the 3-D geometry of it.

Observations at the western limb (Figs. 3, 4, movie\_STB\_27,
movie\_SDO\_04) show bright plasma moving from the south to the
north along high loops above the body of the filament F2. The
height of the top of the loops is about 140 Mm and the height of
the prominence top is about 70 Mm. Since on-disc observations
(Figs. 2, 5, movie\_SDO\_27, movie\_STA\_04) show at the same time
plasma propagating along the inverted-'S'-shaped axis of the
interconnecting flux rope from its south-eastern end to the
north-western end, the motion corresponds to a field-aligned
motion within a left-hand helix. After each plasma injection into
the flux rope at its south-eastern end, some bright and dark
features move back demonstrating at times counterstreaming. Such
backward motion can be recognized in {\it STEREO A}/SECCHI images
on disc in the 195-\AA \ channel on 2012 August 04 between 06:45
UT and 07:05 UT. The backward motion within a left-hand helix
corresponds to the clockwise rotation observed by the {\it
SDO}/AIA at the limb (Fig. 8(a), (b), movie\_SDO\_04).

More detailed examination of a sequence of images shows that the
threads deviating clockwise from the filament axis (Fig. 7) are
located at the bottom side of the flux rope. There are no clear
cases of their overlaying the threads that deviate
counterclockwise, while opposite examples can be easily found
(Fig. 11). Nevertheless, most convincing arguments proving the
left-handedness of the flux rope are in the comparison of plasma
motion in two projections. During a 'westward' motion (from the
south-western end of the filament F1 to the western end of the
filament F2), observations at the limb show a counterclockwise
rotation, while a 'eastward' motion corresponds to a clockwise
rotation. A nearby sunspot visible at the right-hand bottom corner
of Fig. 11 also demonstrates negative helicity. Coronal loops
connected to the sunspot compose a pattern of a counterclockwise
whirl. Such counterclockwise whirls are more typical for the
northern hemisphere but here this is an exception together with
the dextral chirality of the filaments. \citet{b64} found that
filaments with an end curving toward sunspots with
counterclockwise/clockwise whirls are always dextral/sinistral.
Although this sunspot is not connected to filament ends, it shows
the dominating helicity of the region.

In addition to the obvious field-aligned motion, the whole system
of thin threads moves in the direction to the center of the saddle
structure between 16:45 UT and 17 30 UT on July 27, then returns
to the previous position on the periphery of the saddle by 17:52
UT, and makes new rush to the center between 16:53 UT and 18 11
UT. The velocity of the field-aligned motion of different features
is about 100 $\pm$ 10 km s$^{-1}$. The velocity of the transversal
motion of threads during their first rush in the direction to the
center of the saddle is 70 $\pm$ 2 km s$^{-1}$. Rapid
displacements of entire long threads mean rapid changes of the
magnetic field possibly caused by some instability (kink, torus).
At the beginning of several activations, a displacement of
counterclockwise deviated threads from the left side of the flux
rope to the right side is observed (for example, between 14:20 UT
and 14: 57 UT on July 27, see movie\_SDO\_27). This motion
corresponds to untwisting of a left-hand helix and such untwisting
is observed in many eruptive events.

Fig. 12 presents a scheme of the magnetic structure of the two
filaments under study. We assume that each of the filaments is
contained in a left-handed flux rope according to their dextral
chirality. Field lines f1 and f5 in Fig 12 represent these flux
ropes. There is also a flux rope represented by field lines f2,
f3, and f4 connecting the filaments F1 and F2. This flux rope is
located on the PIL turning to the west nearly at a right angle
close to the saddle as shown in Fig. 10. The fine structure and
internal field-aligned plasma motions show also the negative
helicity of the flux rope. It is not loaded with cold dense plasma
and not visible as a filament in H$\alpha$ images. However, it is
revealed many times in EUV at the time of filament activations
during the passage of the region through the solar disc, which
proves its permanent existence. The interconnecting flux rope can
be composed by a number of strands of the flux ropes containing
the main bodies of the filaments F1 and F2, which split in the
vicinity of a null point at the center of the saddle, or be the
separate flux rope like in double-decker filaments. The field line
f4 represents the threads that propagate to the south of the
filament F2 axis around 09 UT on August 04 (Fig. 4(g) - (i), Fig.
5(f) - (i), movie\_SDO\_04). The field line f2 represents the
threads making up the loop shown in Fig. 7(c), (d). This line
approaches closest to the null point and is most curved, as
pointed by the arrow in Fig. 11(c), because of the sharp changes
in the field line geometry near the null point.

The scheme in Fig. 12 is derived from the observed fine structure
of the interacting filaments. However, it is very similar to the
magnetic configuration being formed after two flux ropes
reconnection in the MHD simulation of \citet{b67a}. Due to the
continuous changes of the boundary conditions in the simulations,
the configuration with four braided flux ropes exists only a
rather short period of time. In our case, the configuration is not
as symmetric as in the simulation. Only one interconnecting flux
rope is observed and it exists a long time, possibly, because of
the slow changes in the surrounding photospheric magnetic fields.

\section[]{Summary and Conclusions}

Two filaments located not far from each other at a middle latitude
in the southern hemisphere look to be separate and independent in
H$\alpha$  images in the end of July and in the beginning of
August 2012. The chirality of both filaments is dextral in
violation of the general hemispheric rule. The filaments do not
change significantly their shapes during the passage through the
solar disc. During several pulses of filament activations observed
between July 27 and August 04, a flux rope connecting both
filaments becomes visible in EUV channels. Numerous appearances
and disappearances of the connecting flux rope at the same place
and with the same structure in EUV observations show that this
magnetic flux rope exists permanently but becomes visible only
during activations when it is filled with emitting and absorbing
plasma. However, the fine structure of the flux rope in EUV images
is rather tangled. At first glance it is difficult to determine
even the helicity sign of the flux rope, or handedness of the
observed helix, because it is not so easy to decide to what side
of the flux rope, frontal or back, visible features belong
\citep{b32, b22}. Only the advantage of multi-view-point
observations provided by the {\it SDO} and {\it STEREO} missions
and observations of plasma motions within the flux rope gives the
opportunity to determine reliably the true sign (negative) of
helicity. This sign of the flux-rope helicity corresponds to the
dextral chirality of both filaments.

As is well known, flux ropes are in horizontal equilibrium when
they follow PILs of the ambient magnetic field. Each filament, F1
and F2, is embedded into its own flux rope, which follows its own
PIL (see Fig. 9). The interconnecting flux rope should also follow
a PIL. Therefore, the PIL should exist that connects two filament
channels at some height. Such change in PIL connectivity is
possible only near a null point. EUV observations really show the
presence of a saddle structure at the place of the closest
approachment of the filaments. The center of the saddle structure
is a two-dimensional X-point. The horizontal field vanishes there
owing to the symmetry of the structure. If a PIL, at which the
vertical field is zero, crosses the saddle, this means the
existence of a 3-D null point.

Our potential magnetic field calculations have very limited
accuracy because of the characteristic of the photospheric
magnetic field distribution in the region. Choosing a limited
photospheric area as the boundary condition, we can obtain PILs
connecting the two filament channels (Fig. 10). However, the PIL
connectivity depends on the size of the chosen area. Anyway, these
calculations show the possibility of the existence of the PIL
connecting the two filament channels, which is manifested more
exactly by the presence of the stable flux rope.

The filament activation includes the heating of the plasma and
field-aligned motions within the filament body and the enveloping
flux rope. The velocity of the field-aligned motion of different
features is about 100 $\pm$ 10 km s$^{-1}$. At some moments, a
section of the whole interconnecting flux rope near the saddle
moves towards the null point with a velocity of 70 $\pm$ 2 km
s$^{-1}$. Then it returns back. Sometimes, the rotation of the
flux-rope threads corresponding to its unwinding is observed.

The considered example of filament interaction demonstrates the
complex magnetic structure of the region, which is not evident in
H$\alpha$  observations. It consists of three braided flux ropes
in the vicinity of the coronal null point. The observations,
despite the tangled fine structure in some EUV images, support
flux rope filament models. They give more evidence for the
one-to-one relationship between the filament chirality and the
flux rope helicity.

\section*{Acknowledgments}

The author acknowledges the Big Bear Solar Observatory,
Observatory Paris-Meudon, {\it STEREO}, {\it SDO} teams for the
high-quality data supplied. The author thanks the referee for
critical comments and useful suggestions. Movies created using the
ESA and NASA funded Helioviewer Project. This work was supported
in part by the Russian Foundation for Basic Research (grant
14-02-92690).

\bsp

\label{lastpage}


\begin{thebibliography}{99}

\bibitem[\protect\citeauthoryear{Antiochos}{1998}]{b1}
Antiochos S. K., 1998, ApJ, 502, L181
\bibitem[\protect\citeauthoryear{Antiochos, Dahlburg \& Klimchuk}{Antiochos et al.}{1994}]{b2}
Antiochos S. K., Dahlburg R. B., Klimchuk J. A., 1994, ApJ, 420,
L41
\bibitem[\protect\citeauthoryear{Aulanier, DeVore \& Antiochos}{Aulanier et al.} {2002}]{b3}
Aulanier G., DeVore C. R., Antiochos S. K., 2002, ApJ, 567, L97
\bibitem[\protect\citeauthoryear{Babcock \& Babcock}{1955}]{b4}
Babcock H. W., Babcock H. D., 1955, ApJ, 121, 349
\bibitem[\protect\citeauthoryear{Bateman}{1978}]{b5}
Bateman G., 1978, MHD Instabilities, Massachusetts Institute of
Technology, Cambridge, MA.
\bibitem[\protect\citeauthoryear{Casini et al.}{2003}]{b6}
Casini R., L\'{o}pez Ariste A., Tomczyk S.,  Lites B. W., 2003,
ApJ, 598, L67
\bibitem[\protect\citeauthoryear{Chae}{2000}]{b7}
Chae J., 2000, ApJ, 540, L115
\bibitem[\protect\citeauthoryear{Chandra et al.}{2010}]{b8}
Chandra R., Pariat E., Schmieder B., Mandrini C. H., Uddin W.,
2010, Sol. Phys., 261, 127
\bibitem[\protect\citeauthoryear{Chandra et al.}{2011}]{b8a}
Chandra R., Schmieder B., Mandrini C. H., D$\acute{e}$moulin P.,
T$\ddot{o}$r$\ddot{o}$k T., Pariat E., Uddin W., 2011, Sol. Phys.,
269, 83
\bibitem[\protect\citeauthoryear{Cheng et al.}{2011}]{b9}
Cheng X., Zhang J., Liu Y., Ding M. D., 2011, ApJ, 732, L25
\bibitem[\protect\citeauthoryear{Cheng et al.}{2012}]{b10}
Cheng X., Zhang J., Saar S. H., Ding M. D., 2012, ApJ, 761, 62
\bibitem[\protect\citeauthoryear{Collados, Trujillo Bueno \& Asensio Ramos}{Collados et
al.}{2003}]{b11}
Collados M., Trujillo Bueno J., Asensio Ramos A., 2003, in
Trujillo Bueno J.,  S\'{a}nchez Almeida J., eds, ASP Conf. Ser.
236, Solar Polarization Workshop 3,   Astron. Soc. Pacific, San
Francisco, p. 468
\bibitem[\protect\citeauthoryear{ D$\acute{e}$moulin \& Aulanier}{2010}]{b12}
D$\acute{e}$moulin P., Aulanier G., 2010, ApJ, 718, 1388
\bibitem[\protect\citeauthoryear{Deng et al.}{2002}]{b12a}
Deng Y., Lin Y., Schmieder B., Engvold O., 2002, Sol. Phys. 209,
153
\bibitem[\protect\citeauthoryear{DeVore \& Antiochos}{2000}]{b13}
DeVore C. R., Antiochos S. K., 2000, ApJ, 539, 954
\bibitem[\protect\citeauthoryear{Durrant}{2002}]{b14}
Durrant C. J.,  2002, Sol. Phys., 211, 83
\bibitem[\protect\citeauthoryear{Fan \& Gibson}{2007}]{b15}
Fan Y., Gibson S. E., 2007, ApJ, 668, 1232
\bibitem[\protect\citeauthoryear{Filippov}{2011}]{b16}
Filippov B. P., 2011, Astron. Rep., 55, 541
\bibitem[\protect\citeauthoryear{Filippov}{ 2013a}]{b17}
Filippov B.,  2013a, in Schmieder B.,  Malherbe J.-M.,  Wu S. T.,
eds, Proc. IAU Symp. 300, Nature of Prominences and their Role in
Space Weather, Cambridge University Press, Cambridge, UK, p. 412
\bibitem[\protect\citeauthoryear{Filippov}{2013b}]{b18}
Filippov B., 2013b, ApJ, 773, 10
\bibitem[\protect\citeauthoryear{Filippov \& Den}{2000}]{b19}
Filippov B.P., Den O.G., 2000,  Astron. Lett., 26, 322
\bibitem[\protect\citeauthoryear{Filippov \& Den}{2001}]{b20}
Filippov B.P., Den O.G., 2001, J. Geophys. Res., 106, 25177
\bibitem[\protect\citeauthoryear{Filippov \& Srivastava}{2011}]{b21}
Filippov B., Srivastava A. K., 2011, Sol. Phys., 270, 151
\bibitem[\protect\citeauthoryear{Filippov et al.}{2015}]{b22}
Filippov B., Srivastava A. K., Dwivedi B. N., Masson S., Aulanier
G., Joshi N. C., Uddin W., 2015, MNRAS, 451, 5636
\bibitem[\protect\citeauthoryear{Forbes \& Isenberg}{1991}]{b23}
Forbes T.G., Isenberg P.A., 1991, ApJ, 373, 294
\bibitem[\protect\citeauthoryear{Gosling, Birn \& Hesse}{Gosling et al.}{1995}]{b24}
Gosling J. T., Birn J., Hesse M., 1995, Geophys. Res. Lett., 22,
869
\bibitem[\protect\citeauthoryear{Green, Kliem \& Wallace}{Green et al.}{2011}]{b25}
Green L. M., Kliem B.,  Wallace A. J., 2011, A\&A, 526, A2
\bibitem[\protect\citeauthoryear{Howard \& Harvey}{1964}]{b26}
Howard R. F., Harvey J. W., 1964, ApJ, 139, 1328
\bibitem[\protect\citeauthoryear{Howard et al.}{2008}]{b27}
Howard R.A. et al., 2008, Space Sci. Rev., 136, 67
\bibitem[\protect\citeauthoryear{Ipson et al.}{2005}]{b28}
Ipson S. S., Zharkova V. V., Zharkov S., Benkhalil A. K.,
Aboudarham J., Fuller N., 2005, Sol. Phys., 228, 399
\bibitem[\protect\citeauthoryear{Isobe, Tripathi \& Archontis}{Isobe et al.}{2007}]{b30}
Isobe, H., Tripathi, D., Archontis, V., 2007, ApJ, 657, L53
\bibitem[\protect\citeauthoryear{Isenberg \& Forbes}{2007}]{b31}
Isenberg P. A., Forbes T. G., 2007, ApJ, 670, 1453
\bibitem[\protect\citeauthoryear{Joshi et al.}{2014}]{b32}
Joshi N.C., Srivastava A.K., Filippov B., Kayshap P., Uddin W.,
Chandra R., Choudhary D.P., Dwivedi B.N., 2014, ApJ, 787, 11
\bibitem[\protect\citeauthoryear{Karpen et al.}{2003}]{b33}
Karpen J. T., Antiochos S. K., Klimchuk J. A., MacNeice P. J.,
2003, ApJ, 593, 1187
\bibitem[\protect\citeauthoryear{Kliem \& T$\ddot{o}$r$\ddot{o}$k}{2006}]{b34}
Kliem B.,  T$\ddot{o}$r$\ddot{o}$k T., 2006, Phys. Rev. Lett.,
96(25), 255002
\bibitem[\protect\citeauthoryear{Kliem et al.}{2014}]{b35}
Kliem B.,  T$\ddot{o}$r$\ddot{o}$k T., Titov V. S., Lionello R.,
Linker J. A., Liu R., Liu C., Wang H., 2014, ApJ,  792, 107
\bibitem[\protect\citeauthoryear{Kumar, Manoharan \& Uddin}{Kumar et al.}{2010}]{b36}
Kumar P., Manoharan P. K., Uddin W., 2010, ApJ, 710, 1195
\bibitem[\protect\citeauthoryear{Lemen et al.}{2012}]{b37}
Lemen, J. R. et al., 2012, Sol. Phys., 275, 17
\bibitem[\protect\citeauthoryear{Leroy}{1989}]{b38}
Leroy, J. L. 1989, in ASSL 150, Dynamics and Structure of
Quiescent Solar Prominences,  p. 77
\bibitem[\protect\citeauthoryear{Li \& Zhang}{2013a}]{b39}
Li L. P.,  Zhang J., 2013a, A\&A, 552, L11
\bibitem[\protect\citeauthoryear{Li \& Zhang}{2013b}]{b40}
Li T.,  Zhang J., 2013b, ApJ, 770, L25
\bibitem[\protect\citeauthoryear{Li \& Zhang}{2013c}]{b41}
Li T., Zhang J., 2013c, ApJ, 778, L29
\bibitem[\protect\citeauthoryear{Lin et al.}{1998}]{b42}
Lin J., Forbes T.G., Isenberg P.A.,  D$\acute{e}$moulin P., 1998,
ApJ, 504, 1006
\bibitem[\protect\citeauthoryear{Lites}{2005}]{b45}
Lites B. W., 2005, ApJ, 622, 1275
\bibitem[\protect\citeauthoryear{Liu et al.}{2010}]{b43}
Liu R., Liu C., Wang S., Deng N., Wang H., 2010, ApJ, 725, L84
\bibitem[\protect\citeauthoryear{Liu et al.}{2012}]{b44}
Liu, R., Kliem, B., T$\ddot{o}$r$\ddot{o}$k T., Liu C., Titov V.
S., Lionello R., Linker J. A., Wang H., 2012, ApJ, 756, 59
\bibitem[\protect\citeauthoryear{Low}{1996}]{b46}
Low B. C., 1996, Sol. Phys., 167, 217
\bibitem[\protect\citeauthoryear{Mackay, \& van Ballegooijen}{2005}]{b47}
Mackay D. H., van Ballegooijen A. A., 2005, ApJ, 621, L77
\bibitem[\protect\citeauthoryear{Mackay et al.}{2010}]{b48}
Mackay D. H., Karpen J. T., Ballester J. L., Schmieder B.,
Aulanier G., 2010, Space Sci. Rev., 151, 333
\bibitem[\protect\citeauthoryear{Martens \& Zwaan}{2001}]{b49}
Martens P. C.,  Zwaan, C., 2001, ApJ, 558, 872
\bibitem[\protect\citeauthoryear{Martin}{1998a}]{b50}
Martin S. F., 1998a,  Sol. Phys., 182, 107
\bibitem[\protect\citeauthoryear{Martin}{1998b}]{b51}
Martin S. F., 1998b, in  Webb D.,  Rust D., Schmieder B., eds, IAU
Colloquium 167, ASP Conference Series 150, New Perspectives on
Solar Prominences,  p. 419
\bibitem[\protect\citeauthoryear{Martin, Bilimoria \& Tracadas}{Martin et al.}{1994}]{b52}
Martin S. F., Bilimoria R., Tracadas P. W., 1994, in Solar Surface
Magnetism,  Kluwer, Dordrecht, p. 303
\bibitem[\protect\citeauthoryear{McIntosh}{1972}]{b53}
McIntosh P. S., 1972, Rev. Geophys. Space Phys., 10, 837
\bibitem[\protect\citeauthoryear{McKenzie \& Canfield}{2008}]{b54}
McKenzie D. E., Canfield R. C., 2008, A\&A, 481, L65
\bibitem[\protect\citeauthoryear{Nindos, Patsourakos \& Wiegelmann}{Nindos et al.}{2012}]{b55}
Nindos A., Patsourakos S., Wiegelmann T., 2012, ApJL 748, L6
\bibitem[\protect\citeauthoryear{Okamoto et al.}{2008}]{b56}
Okamoto T. J., Tsuneta S., Lites B. W., et al., 2008, ApJ, 673,
L215
\bibitem[\protect\citeauthoryear{Olmedo \& Zhang}{2010}]{b57}
Olmedo O., Zhang J., 2010, ApJ, 718, 433
\bibitem[\protect\citeauthoryear{Patsourakos, Vourlidas \& Stenborg}{Patsourakos et al.}{2013}]{b58}
Patsourakos S., Vourlidas A., Stenborg G., 2013, ApJ, 764, 125
\bibitem[\protect\citeauthoryear{Pevtsov, Balasubramaniam \& Roger}{Pevtsov et al.}{2003}]{b59}
Pevtsov A. A., Balasubramaniam K. S., Rogers J. W., 2003, ApJ,
595, 500
\bibitem[\protect\citeauthoryear{Pneuman}{1983}]{b60}
Pneuman G. W., 1983, Sol. Phys., 88, 219
\bibitem[\protect\citeauthoryear{Raouafi}{2009}]{b61}
Raouafi N.-E., 2009, ApJ, 691, L128
\bibitem[\protect\citeauthoryear{Rust}{1999}]{b62}
Rust D. M., 1999, in Brown M. R., Canfield R. C., Pevtsov A. A.,
eds, Geophysical Monograph 111, Magnetic Helicity in Space and
Laboratory Plasmas, American Geophysical Union, Washington, p. 221
\bibitem[\protect\citeauthoryear{Rust \& Kumar}{1994}]{b63}
Rust D. M., Kumar A., 1994, Sol. Phys., 155, 69
\bibitem[\protect\citeauthoryear{Rust \& Martin}{1994}]{b64}
Rust D. M.,  Martin S. F., 1994, in Balasubramanian K. S., Simon
G.W., eds, ASP Conf. Ser. 68, Solar Active Region Evolution:
Comparing Models with Observations, p. 337
\bibitem[\protect\citeauthoryear{Schmieder et al.}{2004}]{b64a}
Schmieder B., Mein  N., Deng Y., Dumitrache C. Malherbe J. M.,
Staiger J., DeLuca  E. E., 2004, Sol. Phys., 223, 119
\bibitem[\protect\citeauthoryear{Schou et al.}{2012}]{b65}
Schou J., et al., 2012, Sol. Phys., 275, 229
\bibitem[\protect\citeauthoryear{Smith \& Ramsey}{1967}]{b66}
Smith S. F., Ramsey H. E., 1967, Sol. Phys., 2, 158
\bibitem[\protect\citeauthoryear{Snodgrass, Kress \& Wilson}{Snodgrass et
al.}{2000}]{b67}
Snodgrass H. B., Kress J. M., Wilson P. R., 2000,
Sol. Phys., 191, 1
\bibitem[\protect\citeauthoryear{ T$\ddot{o}$r$\ddot{o}$k et al.}{2011}]{b67a}
T$\ddot{o}$r$\ddot{o}$k T., Chandra R., Pariat E.,
D$\acute{e}$moulin P., Schmieder B., Aulanier G., Linton M. G.,
Mandrini C. H., 2011, ApJ, 728, 65
\bibitem[\protect\citeauthoryear{van Ballegooijen \& Martens}{1989}]{b68}
van Ballegooijen A. A.,  Martens, P. C. H., 1989, ApJ, 343, 971
\bibitem[\protect\citeauthoryear{van Tend \& Kuperus}{1978}]{b69}
van Tend W., Kuperus M., 1978, Sol. Phys., 59, 115
\bibitem[\protect\citeauthoryear{Wuelser et al.}{2004}]{b70}
Wuelser J.-P. et al., 2004, Proc. SPIE, 5171, p. 111
\bibitem[\protect\citeauthoryear{Zhang \& Low}{2005}]{b71}
Zhang M., Low, B. C., 2005, ARA\&A, 43, 103
\bibitem[\protect\citeauthoryear{Zhang, Cheng \& Ding}{Zhang et al.}{2012}]{b72}
Zhang J., Cheng X., Ding M.-D., 2012, NatCo, 3, 747



\end{thebibliography}
\end{document}